\documentclass[twocolumn,amsmath,amssymb,groupedaddress,prl]{revtex4}

\usepackage{epsfig}

\def\al{\alpha}
\def\be{\beta}
\def\ga{\gamma}
\def\de{\delta}

\def\ka{\kappa}
\def\la{\lambda}

\def\si{\sigma}

\def\ta{\tau}

\def\vp{\varphi}
\def\ch{\chi}

\def\om{\omega}

\def\De{\Delta}

\def\fr#1#2{{{#1} \over {#2}}}

\def\frac#1#2{{\textstyle{{#1}\over {#2}}}}

\def\lsim{\mathrel{\rlap{\lower4pt\hbox{\hskip1pt$\sim$}}
    \raise1pt\hbox{$<$}}}
\def\gsim{\mathrel{\rlap{\lower4pt\hbox{\hskip1pt$\sim$}}
    \raise1pt\hbox{$>$}}}

\def\etal{{\it et al.}}

\newcommand{\beq}{\begin{equation}}
\newcommand{\eeq}{\end{equation}}
\newcommand{\bea}{\begin{eqnarray}}
\newcommand{\eea}{\end{eqnarray}}
\newcommand{\bit}{\begin{itemize}}
\newcommand{\eit}{\end{itemize}}
\newcommand{\rf}[1]{(\ref{#1})}

\def\kddr{{(k^{(6)}_1)}{}}
\def\krr{{(k^{(6)}_2)}{}}

\def\kb{\overline{k}{}}

\def\kbddr{{(\kb^{(6)}_1)}{}}
\def\kbrr{{(\kb^{(6)}_2)}{}}

\def\kl{{\ka\la}}

\def\abgd{{\al\be\ga\de}}

\def\klmn{{\ka\la\mu\nu}}
\def\ol#1{\overline{#1}}

\newcommand{\kei}[1]{(\kb_{\rm eff})_{#1}}

\begin{document}

\title{Combined search for Lorentz violation in short-range gravity}

\author{Cheng-Gang Shao}
\author{Yu-Jie Tan}
\author{Wen-Hai Tan}
\author{Shan-Qing Yang}
\author{Jun Luo}
\affiliation{
MOE Key Laboratory of Fundamental Physical Quantities Measurements,
School of Physics, Huazhong University of Science and Technology, 
Wuhan 430074, 
People's Republic of China}

\author{Michael Edmund Tobar}
\affiliation{School of Physics, University of Western Australia, 
Crawley, WA 6009, Australia}

\author{Quentin G.\ Bailey}
\affiliation{
Physics Department, Embry-Riddle Aeronautical University,
Prescott, AZ 86301, U.S.A.}

\author{J.C.\ Long}
\author{E.\ Weisman}
\author{Rui Xu}
\author{V.\ Alan Kosteleck\'y}
\affiliation{Physics Department, Indiana University, 
Bloomington, Indiana 47405, U.S.A.}

\date{March 2016;
accepted for publication in {\it Physical Review Letters}} 

\begin{abstract}

Short-range experiments testing the gravitational inverse-square law 
at the submillimeter scale
offer uniquely sensitive probes of Lorentz invariance.
A combined analysis of results from the short-range gravity experiments 
HUST-2015, HUST-2011, IU-2012, and IU-2002 
permits the first independent measurements
of the 14 nonrelativistic coefficients for Lorentz violation
in the pure-gravity sector at the level of $10^{-9}$ m$^2$,
improving by an order of magnitude
the sensitivity to numerous types of Lorentz violation 
involving quadratic curvature derivatives and curvature couplings.
 
\end{abstract}

\maketitle

General relativity offers an impressive 
description of gravity at the classical level.
A key ingredient in its construction  
is local Lorentz invariance,
which insures rotation and boost symmetry in a freely falling frame. 
However,
achieving a consistent unification
of gravity with quantum physics
may require modifications of the foundations of general relativity.
These modifications could induce observable violations
of Lorentz invariance,
arising in an underlying unified theory such as strings 
\cite{ksp}.
Experimental tests of Lorentz symmetry in gravity 
therefore have the potential to offer insight
about the structure of physics beyond general relativity
\cite{tables,lvgrreview}.

One interesting option for investigating 
Lorentz violation in pure gravity
is offered by short-range experiments at the submillimeter scale
\cite{review}.
Effective field theory for Lorentz violation in gravity
\cite{akgrav}
provides a generic and model-independent approach 
to studying possible experimental signals of Lorentz violation.
Applying this method reveals 
that quadratic curvature derivatives and quadratic curvature couplings
can produce novel signals in experiments on short-range gravity
\cite{bkx}.
Most of these couplings remain experimentally unexplored
and have {\it a priori} unknown sizes,
with even comparatively large Lorentz violation
remaining viable in certain `countershaded' models
\cite{kt09},
so direct searches without preconceived notions of sensitivity are essential.

Short-range experiments 
at Indiana University (IU) \cite{lk} 
and Huazhong University of Science and Technology (HUST) \cite{hust15}
have achieved sensitivities 
at the level of $10^{-7}$ to $10^{-8}$ m$^{2}$ 
to individual coefficients controlling 
these types of gravitational local Lorentz violation.
However, 
any one short-range experiment measures only nine signal components, 
which is insufficient to constrain simultaneously 
all the predicted effects.
In this work,
we present a combined analysis of results 
from tests of short-range gravity 
based on four different experimental designs
performed
at HUST 
(HUST-2015 \cite{hust15-2} and HUST-2011 \cite{hust15})
and at IU 
(IU-2012 and IU-2002 \cite{lk}).
Our analysis yields the first independent measures 
of all 14 accessible coefficients for Lorentz violation,
consistent with no effect at the level of $10^{-9}$ m$^{2}$.

The Lorentz-violating 
quadratic curvature derivatives and couplings
produce a perturbative correction
to the Newton gravitational potential between two test masses $m_1$, $m_2$
that is inverse cubic and varies with orientation and sidereal time $T$
\cite{bkx}.
Its explicit form is
\beq
V_{\rm LV} (\vec{r}) 
= - G_N \fr{{m_1}{m_2}}{|\vec{r}|^{3}}{\ol k}(\hat{r},T),
\label{pot}
\eeq
where the vector $\vec{r}=\vec{x}_{1}-\vec{x}_{2}$ 
separates $m_1$ and $m_2$,
and 
\beq
{\ol k}=
\frac 32  (\kb_{\rm eff})_{jkjk}
- 9  (\kb_{\rm eff})_{jkll} \hat{r}^j \hat{r}^k 
+ \frac {15}{2} (\kb_{\rm eff})_{jklm} 
\hat{r}^j \hat{r}^k \hat{r}^l \hat{r}^m
\label{k}
\eeq
involves the projection $\hat{r}^{j}$ of the unit vector 
along $\vec{r}$ in the $j$th direction.
The nonrelativistic coefficients $(\kb_{\rm eff})_{jklm}$ 
for Lorentz violation
have dimensions of squared length
and are totally symmetric,
thus containing 15 independent degrees of freedom.
However,
the rotation invariant $(\kb_{\rm eff})_{jkjk}$
produces only a contact correction to the usual Newton force,
so only 14 of them are independently measurable
in short-range experiments.

The coefficients $(\kb_{\rm eff})_{jklm}$ 
take different forms in different inertial frames,
so a canonical Sun-centered celestial-equatorial frame
\cite{tables,sunframe}
is conventionally adopted to report experimental results,
with the $Z$ axis along the direction of the Earth's rotation
and the $X$ axis pointing to the vernal equinox.
The coefficients can be taken constant in this frame 
\cite{sme},
but the rotation of the Earth 
implies that the coefficients in a laboratory frame change with time
and therefore produce sidereal signals in experimental data
\cite{akmeson}.
Neglecting the Earth's boost $\be_\oplus\simeq 10^{-4}$, 
the conversion from the Sun-centered frame $(X, Y, Z)$ 
to a laboratory frame $(x, y, z)$
with $x$ axis pointing to local south
and $z$ axis to the zenith 
can be implemented by the time-dependent rotation 
\beq
R^{jJ}=\left(
\begin{array}{ccc}
\cos\ch\cos\om_\oplus T
&
\cos\ch\sin\om_\oplus T
&
-\sin\ch
\\
-\sin\om_\oplus T
&
\cos\om_\oplus T
&
0
\\
\sin\ch\cos\om_\oplus T
&
\sin\ch\sin\om_\oplus T
&
\cos\ch
\end{array}
\right),
\label{rotmat}
\eeq
where
$\om_\oplus\simeq 2\pi/(23{\rm ~h} ~56{\rm ~min})$ 
is the sidereal frequency.
The laboratory colatitude $\ch$ is 
$\ch\simeq 1.038$ rad for HUST-2015 and HUST-2011,
$\ch\simeq 0.887$ rad for IU-2012,
and $\ch\simeq 0.872$ rad for IU-2002.
The relation between the laboratory coefficients
$(\kb_{\rm eff})_{jklm}$
and the coefficients $(\kb_{\rm eff})_{JKLM}$
in the Sun-centered frame
is therefore
\beq
(\kb_{\rm eff})_{jklm} 
= R^{jJ} R^{kK} R^{lL} R^{mM} (\kb_{\rm eff})_{JKLM}.
\label{rot}
\eeq
It follows that the time oscillations
of the inverse-cube potential \rf{pot}
contain harmonic frequencies up to $4\om_\oplus$.

Most experimental tests of the gravitational inverse-square law 
adopt a planar test-mass geometry
to reduce conventional effects from Newton gravity. 
The Lorentz-violating force between two parallel plates 
can be calculated by numerical integration. 
Since only four harmonic frequencies appear,
the signal from any one short-range experiment 
can contain at most nine Fourier components,
including the DC response.
Two or more experiments are therefore required
to measure simultaneously 
all 14 independent Lorentz-violating degrees of freedom.
Here,
we achieve this using data 
from the recent experiment HUST-2015 \cite{hust15-2}
and from the earlier experiments 
IU-2012 \cite{lk}, HUST-2011 \cite{hust15},
and IU-2002
\cite{jcl03}.
Since the relevant methodologies for the latter three  
are described in detail elsewhere
\cite{lk,jcl02,jcl03,jcl14,hust15},
we focus here on the corresponding analysis for HUST-2015.

The basic design and the operation of the experiment HUST-2015
are described in Ref.\ \cite{hust15-2}.
A bilaterally symmetric I-shaped pendulum
is suspended next to an attractor disk with eightfold symmetry.
The pendulum contains two pure-tungsten test masses,
together with two additional tungsten pieces
designed to compensate the Newton gravitational force 
at the signal frequency. 
The attractor disk consists of eight tungsten source masses
and eight compensation masses.
The centers of the attractor disk and the torsion pendulum are aligned,
and the distance between the surfaces of the test and source masses 
is maintained at 295 $\mu$m.
The pendulum twist is controlled 
using a feedback technique,
by applying differential voltages 
to the two capacitive actuators on the pendulum. 
In the presence of Lorentz violation,
rotating the attractor disk generates a torque.
The signal and disturbance frequencies are well separated, 
so a high measurement resolution can be achieved. 
The apparatus is designed to produce approximate null measurements 
by double compensating for both the test and the source masses. 
When the attractor disk 
rotates at frequency $f_0=2\pi/$(3846.12 s), 
the nominal signal torque oscillates at frequency $8f_0$
The torque is maximal when the source and test masses are face to face
and minimal when they are offset.
However, 
the Lorentz-violating force 
between two finite flat plates is dominated by edge effects 
\cite{hust15}, 
so the Lorentz-violating torque oscillates 
primarily at the frequency $16f_0$
and is an order of magnitude larger than the signal at $8f_0$. 
The resolution of the pendulum is calibrated gravitationally
by rotating a nearby copper cylinder at frequency $f_c=2\pi/$(400 s),
chosen to be close to $8f_0 \simeq 2\pi/$(481 s)
but sufficiently offset to be readily distinguished from it.

\begin{figure}
\includegraphics[width=\hsize]{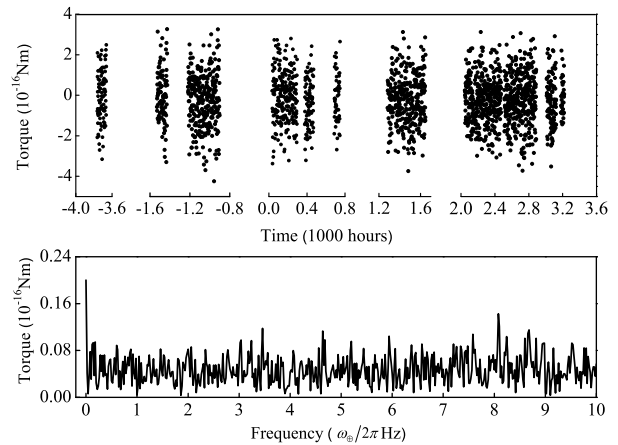}
\vskip -10pt
\caption{
HUST-2015 data at $16 f_0$ and Fourier transform.
\label{dat}}
\end{figure}
 
Data were acquired from December 2014 to August 2015,
during a period of over 2000 hours. 
To extract the signal, 
the recorded data were separated into slices 
according to the modulation period, 
$\De T =3846.12$ s. 
For each slice, 
the Lorentz-violating torque $\ta_{\rm LV}$ was extracted
by fitting the measured torque $\ta^z(T)$ as 
\beq
\ta ^z(T) = {\ta _{\rm LV}}(T) \cos (32\pi {f_0}T + \vp),
\eeq
where the initial phase $\vp$ 
is set by the operation of the experiment. 
We take $\ta_{\rm LV}(T)$ as approximately constant in each data slice
because $\om_{\oplus}\De T\ll 1$ 
and so any sidereal variation within each $\De T$ is negligible.
The upper panel of Fig.\ \ref{dat} displays the extracted torque 
${\ta_{LV}}$ 
as a function of time.
Each point represents the mean of the measurement in $\De T$ without error, 
which is dominated by statistical uncertainty. 
In the Sun-centered frame,
the time origin $T=0$ is defined as the vernal equinox 2000.
For the analysis,
it suffices to use a convenient shifted time $T_\oplus$ 
with origin set when local east coincides 
with the $Y$ axis of the Sun-centered frame
\cite{sunframe}.
The lower panel of Fig.\ \ref{dat}
shows the Fourier spectrum of the torque.

\renewcommand{\tabcolsep}{3pt}
\begin{table} 
\begin{tabular}{ccccc}
\hline
\hline
Mode & HUST-2015 & HUST-2011 & IU-2012 & IU-2002 \\
\hline
$C_{0}$	&	$	-0.20	\pm	2.40	$	&	$	-0.22	\pm	1.90	$	&	$	-31	\pm	120	$	&	$	12	\pm	203	$	\\
$C_{1}$	&	$	0.00	\pm	0.08	$	&	$	0.13	\pm	0.44	$	&	$	-77	\pm	170	$	&	$	34	\pm	123	$	\\
$S_{1}$	&	$	-0.01	\pm	0.08	$	&	$	-0.40	\pm	0.45	$	&	$	-7	\pm	154	$	&	$	-98	\pm	242	$	\\
$C_{2}$	&	$	-0.01	\pm	0.08	$	&	$	-0.04	\pm	0.45	$	&	$	16	\pm	154	$	&	$	-66	\pm	278	$	\\
$S_{2}$	&	$	-0.09	\pm	0.08	$	&	$	0.20	\pm	0.45	$	&	$	-151	\pm	167	$	&	$	-52	\pm	139	$	\\
$C_{3}$	&	$	0.01	\pm	0.08	$	&	$	-0.30	\pm	0.45	$	&	$	-164	\pm	144	$	&	$	207	\pm	141	$	\\
$S_{3}$	&	$	-0.06	\pm	0.08	$	&	$	0.25	\pm	0.45	$	&	$	181	\pm	176	$	&	$	-144	\pm	216	$	\\
$C_{4}$	&	$	0.04	\pm	0.08	$	&	$	-0.06	\pm	0.45	$	&	$	2	\pm	142	$	&	$	26	\pm	223	$	\\
$S_{4}$	&	$	-0.03	\pm	0.08	$	&	$	0.05	\pm	0.45	$	&	$	-10	\pm	165	$	&	$	74	\pm	159	$	\\
\hline
\hline
\end{tabular}
\caption{\label{tab1}  
Fourier amplitudes 
(2$\si$, units $10^{-16}$ Nm for HUST 
and  $10^{-16}$ N for IU).}
\end{table}

\renewcommand{\tabcolsep}{10pt}
\begin{table}
\begin{tabular}{cc}
\hline
\hline
Coefficient & Measurement \\
\hline
$	\kei{XXXX}	$&$	6.4	\pm	32.9	$	\\
$	\kei{XXXY}	$&$	0.0	\pm	8.1	$	\\
$	\kei{XXXZ}	$&$	-2.0	\pm	2.6	$	\\
$	\kei{XXYY}	$&$	-0.9	\pm	10.9	$	\\
$	\kei{XXYZ}	$&$	1.1	\pm	1.2	$	\\
$	\kei{XXZZ}	$&$	-2.6	\pm	17.1	$	\\
$	\kei{XYYY}	$&$	3.9	\pm	8.1	$	\\
$	\kei{XYYZ}	$&$	-0.6	\pm	1.2	$	\\
$	\kei{XYZZ}	$&$	-1.0	\pm	1.0	$	\\
$	\kei{XZZZ}	$&$	-8.1	\pm	10.3	$	\\
$	\kei{YYYY}	$&$	7.0	\pm	32.9	$	\\
$	\kei{YYYZ}	$&$	0.3	\pm	2.6	$	\\
$	\kei{YYZZ}	$&$	-2.5	\pm	17.1	$	\\
$	\kei{YZZZ}	$&$	3.6	\pm	10.2	$	\\
\hline
\hline
\end{tabular}
\caption{\label{combined}
Independent coefficient values 
(2$\si$, units $10^{-9}$~m$^{2}$)
obtained by combining HUST and IU data
\cite{lk,hust15,hust15-2}.}
\end{table}

\renewcommand{\tabcolsep}{5pt}
\begin{table*}
\begin{tabular}{cc|cc|cc}
\hline
\hline
Coefficient & Measurement &
Coefficient & Measurement &
Coefficient & Measurement \\ 
\hline
$	\kbddr_{	XTXTXX	}	$	&	$	-0.8	\pm	6.3	$	&	$	\kbddr_{	XYXYYY	}	$	&	$	-0.2	\pm	36.3	$	&	$	\kbddr_{	XZYZZZ	}	$	&	$	-3.2	\pm	2.9	$	\\
$	\kbddr_{	XTXTXY	}	$	&	$	1.6	\pm	2.3	$	&	$	\kbddr_{	XYXYYZ	}	$	&	$	1.4	\pm	1.8	$	&	$	\kbddr_{	YTYTXX	}	$	&	$	0.1	\pm	9.9	$	\\
$	\kbddr_{	XTXTXZ	}	$	&	$	1.2	\pm	1.2	$	&	$	\kbddr_{	XYXYZZ	}	$	&	$	-1.8	\pm	7.6	$	&	$	\kbddr_{	YTYTXY	}	$	&	$	1.3	\pm	2.3	$	\\
$	\kbddr_{	XTXTYY	}	$	&	$	-0.4	\pm	27.6	$	&	$	\kbddr_{	XYXZXX	}	$	&	$	3.2	\pm	3.7	$	&	$	\kbddr_{	YTYTXZ	}	$	&	$	1.4	\pm	1.5	$	\\
$	\kbddr_{	XTXTYZ	}	$	&	$	-1.2	\pm	1.5	$	&	$	\kbddr_{	XYXZXY	}	$	&	$	0.1	\pm	1.5	$	&	$	\kbddr_{	YTYTYY	}	$	&	$	-3.1	\pm	11.8	$	\\
$	\kbddr_{	XTXTZZ	}	$	&	$	2.2	\pm	12.4	$	&	$	\kbddr_{	XYXZXZ	}	$	&	$	-1.5	\pm	1.5	$	&	$	\kbddr_{	YTYTYZ	}	$	&	$	-0.5	\pm	1.2	$	\\
$	\kbddr_{	XTYTXX	}	$	&	$	0.0	\pm	8.1	$	&	$	\kbddr_{	XYXZYY	}	$	&	$	-1.0	\pm	1.5	$	&	$	\kbddr_{	YTYTZZ	}	$	&	$	2.8	\pm	14.1	$	\\
$	\kbddr_{	XTYTXY	}	$	&	$	-0.1	\pm	6.9	$	&	$	\kbddr_{	XYXZYZ	}	$	&	$	0.8	\pm	4.5	$	&	$	\kbddr_{	YTZTXX	}	$	&	$	-3.2	\pm	3.7	$	\\
$	\kbddr_{	XTYTXZ	}	$	&	$	-1.6	\pm	1.8	$	&	$	\kbddr_{	XYXZZZ	}	$	&	$	-1.4	\pm	1.8	$	&	$	\kbddr_{	YTZTXY	}	$	&	$	0.8	\pm	1.8	$	\\
$	\kbddr_{	XTYTYY	}	$	&	$	-4.0	\pm	8.1	$	&	$	\kbddr_{	XYYZYY	}	$	&	$	1.7	\pm	3.7	$	&	$	\kbddr_{	YTZTXZ	}	$	&	$	1.6	\pm	1.5	$	\\
$	\kbddr_{	XTYTYZ	}	$	&	$	0.8	\pm	1.8	$	&	$	\kbddr_{	XYYZYZ	}	$	&	$	1.5	\pm	1.5	$	&	$	\kbddr_{	YTZTYY	}	$	&	$	-0.3	\pm	2.6	$	\\
$	\kbddr_{	XTYTZZ	}	$	&	$	3.2	\pm	2.9	$	&	$	\kbddr_{	XYYZZZ	}	$	&	$	-1.4	\pm	2.6	$	&	$	\kbddr_{	YTZTZZ	}	$	&	$	-3.6	\pm	10.2	$	\\
$	\kbddr_{	XTZTXX	}	$	&	$	2.0	\pm	2.6	$	&	$	\kbddr_{	XZXZXX	}	$	&	$	-0.5	\pm	27.6	$	&	$	\kbddr_{	YZYZYY	}	$	&	$	-0.5	\pm	27.6	$	\\
$	\kbddr_{	XTZTXY	}	$	&	$	-1.6	\pm	1.8	$	&	$	\kbddr_{	XZXZXY	}	$	&	$	-2.0	\pm	5.7	$	&	$	\kbddr_{	YZYZYZ	}	$	&	$	-3.2	\pm	3.7	$	\\
$	\kbddr_{	XTZTXZ	}	$	&	$	0.1	\pm	9.1	$	&	$	\kbddr_{	XZXZXZ	}	$	&	$	1.7	\pm	3.7	$	&	$	\kbddr_{	ZTZTXX	}	$	&	$	0.2	\pm	9.8	$	\\
$	\kbddr_{	XTZTYY	}	$	&	$	1.7	\pm	3.7	$	&	$	\kbddr_{	XZXZYY	}	$	&	$	-1.7	\pm	6.5	$	&	$	\kbddr_{	ZTZTXY	}	$	&	$	1.4	\pm	1.4	$	\\
$	\kbddr_{	XTZTYZ	}	$	&	$	1.6	\pm	1.5	$	&	$	\kbddr_{	XZXZYZ	}	$	&	$	1.0	\pm	1.5	$	&	$	\kbddr_{	ZTZTXZ	}	$	&	$	2.2	\pm	2.0	$	\\
$	\kbddr_{	XTZTZZ	}	$	&	$	8.1	\pm	10.3	$	&	$	\kbddr_{	XZXZZZ	}	$	&	$	-0.2	\pm	36.3	$	&	$	\kbddr_{	ZTZTYZ	}	$	&	$	-1.3	\pm	2.0	$	\\
$	\kbddr_{	XYXYXY	}	$	&	$	3.2	\pm	2.9	$	&	$	\kbddr_{	XZYZYY	}	$	&	$	2.0	\pm	5.7	$	&	$	\kbddr_{	ZTZTZZ	}	$	&	$	0.6	\pm	10.9	$	\\
$	\kbddr_{	XYXYXZ	}	$	&	$	-1.4	\pm	2.6	$	&	$	\kbddr_{	XZYZYZ	}	$	&	$	0.1	\pm	1.5	$	&						&						\\
\hline
\hline
\end{tabular}
\caption{\label{k1results}
Constraints 
(2$\si$, units $10^{-9}$~m$^{2}$)
on 59 independent coefficients $\kddr_{\abgd\kl}$ 
taken one at a time.}
\end{table*}

\renewcommand{\tabcolsep}{5pt}
\begin{table*}
\begin{tabular}{cc|cc|cc}
\hline
\hline
Coefficient & Measurement &
Coefficient & Measurement &
Coefficient & Measurement \\ 
\hline
$\kbrr_{XTXTXTXT}$	&$	0.7	\pm	3.0	$&	$\kbrr_{XTZTXYXY}$	&$	-0.2	\pm	0.3	$&	$\kbrr_{ZTZTZTZT}$	&$	0.6	\pm	4.3	$	\\
$\kbrr_{XTXTXTYT}$	&$	0.0	\pm	1.0	$&	$\kbrr_{XTZTXYXZ}$	&$	-0.2	\pm	0.2	$&	$\kbrr_{ZTZTXYXZ}$	&$	0.5	\pm	1.3	$	\\
$\kbrr_{XTXTXTZT}$	&$	-0.2	\pm	0.3	$&	$\kbrr_{XTZTXYYZ}$	&$	0.0	\pm	1.1	$&	$\kbrr_{ZTZTXYYZ}$	&$	1.0	\pm	1.3	$	\\
$\kbrr_{XTXTXYXY}$	&$	0.3	\pm	1.4	$&	$\kbrr_{XTZTXZXZ}$	&$	-0.3	\pm	0.3	$&	$\kbrr_{ZTZTXZXZ}$	&$	0.3	\pm	2.1	$	\\
$\kbrr_{XTXTXYXZ}$	&$	0.4	\pm	0.5	$&	$\kbrr_{XTZTXZYZ}$	&$	0.2	\pm	0.2	$&	$\kbrr_{ZTZTXZYZ}$	&$	-0.4	\pm	0.4	$	\\
$\kbrr_{XTXTXYYZ}$	&$	0.2	\pm	0.3	$&	$\kbrr_{XTZTYTYT}$	&$	-0.2	\pm	0.5	$&	$\kbrr_{XYXYXYXY}$	&$	0.4	\pm	2.6	$	\\
$\kbrr_{XTXTXZXZ}$	&$	0.4	\pm	1.5	$&	$\kbrr_{XTZTYTZT}$	&$	-0.2	\pm	0.2	$&	$\kbrr_{XYXYXYXZ}$	&$	0.2	\pm	0.3	$	\\
$\kbrr_{XTXTXZYZ}$	&$	0.0	\pm	1.0	$&	$\kbrr_{XTZTYZYZ}$	&$	-0.3	\pm	0.4	$&	$\kbrr_{XYXYXYYZ}$	&$	0.2	\pm	0.3	$	\\
$\kbrr_{XTXTYTYT}$	&$	0.1	\pm	3.5	$&	$\kbrr_{XTZTZTZT}$	&$	-1.0	\pm	1.3	$&	$\kbrr_{XYXYXZXZ}$	&$	0.4	\pm	2.1	$	\\
$\kbrr_{XTXTYTZT}$	&$	0.4	\pm	0.5	$&	$\kbrr_{YTYTXYXY}$	&$	0.3	\pm	1.4	$&	$\kbrr_{XYXYXZYZ}$	&$	0.2	\pm	0.7	$	\\
$\kbrr_{XTXTYZYZ}$	&$	0.1	\pm	3.5	$&	$\kbrr_{YTYTXYXZ}$	&$	0.0	\pm	0.3	$&	$\kbrr_{XYXYYZYZ}$	&$	0.5	\pm	2.1	$	\\
$\kbrr_{XTXTZTZT}$	&$	0.0	\pm	4.5	$&	$\kbrr_{YTYTXYYZ}$	&$	0.2	\pm	0.5	$&	$\kbrr_{XYXZXYXZ}$	&$	-0.5	\pm	1.7	$	\\
$\kbrr_{XTYTXTYT}$	&$	0.0	\pm	1.7	$&	$\kbrr_{YTYTXZXZ}$	&$	0.1	\pm	3.5	$&	$\kbrr_{XYXZXZXZ}$	&$	0.4	\pm	0.4	$	\\
$\kbrr_{XTYTXTZT}$	&$	0.2	\pm	0.2	$&	$\kbrr_{YTYTXZYZ}$	&$	0.5	\pm	1.0	$&	$\kbrr_{XYXZYZYZ}$	&$	0.1	\pm	0.3	$	\\
$\kbrr_{XTYTXYXY}$	&$	0.2	\pm	0.7	$&	$\kbrr_{YTYTYTYT}$	&$	0.8	\pm	3.0	$&	$\kbrr_{XYYZXYXZ}$	&$	0.2	\pm	0.2	$	\\
$\kbrr_{XTYTXYXZ}$	&$	-0.1	\pm	0.2	$&	$\kbrr_{YTYTYTZT}$	&$	0.0	\pm	0.3	$&	$\kbrr_{XYYZXYYZ}$	&$	-0.3	\pm	1.4	$	\\
$\kbrr_{XTYTXYYZ}$	&$	-0.2	\pm	0.2	$&	$\kbrr_{YTYTYZYZ}$	&$	0.5	\pm	1.7	$&	$\kbrr_{XYYZXZYZ}$	&$	-0.2	\pm	0.2	$	\\
$\kbrr_{XTYTXZXZ}$	&$	-0.3	\pm	0.3	$&	$\kbrr_{YTZTXYXY}$	&$	0.2	\pm	0.3	$&	$\kbrr_{XYYZYZYZ}$	&$	0.3	\pm	0.4	$	\\
$\kbrr_{XTYTXZYZ}$	&$	0.0	\pm	0.9	$&	$\kbrr_{YTZTXYYZ}$	&$	0.2	\pm	0.2	$&	$\kbrr_{XZXZXYYZ}$	&$	0.3	\pm	0.3	$	\\
$\kbrr_{XTYTYTYT}$	&$	0.5	\pm	1.0	$&	$\kbrr_{YTZTXZXZ}$	&$	0.4	\pm	0.4	$&	$\kbrr_{XZXZXZXZ}$	&$	0.0	\pm	2.4	$	\\
$\kbrr_{XTYTYTZT}$	&$	-0.1	\pm	0.2	$&	$\kbrr_{YTZTXZYZ}$	&$	-0.1	\pm	0.2	$&	$\kbrr_{XZXZXZYZ}$	&$	-0.3	\pm	0.3	$	\\
$\kbrr_{XTYTYZYZ}$	&$	-0.3	\pm	0.3	$&	$\kbrr_{YTZTYZYZ}$	&$	0.1	\pm	0.3	$&	$\kbrr_{XZYZXYXZ}$	&$	-0.1	\pm	0.2	$	\\
$\kbrr_{XTYTZTZT}$	&$	-0.4	\pm	0.4	$&	$\kbrr_{YTZTZTZT}$	&$	0.5	\pm	1.3	$&	$\kbrr_{XZYZXZYZ}$	&$	-0.2	\pm	1.1	$	\\
$\kbrr_{XTZTXTZT}$	&$	0.0	\pm	2.3	$&	$\kbrr_{ZTZTYZYZ}$	&$	0.6	\pm	2.9	$&	$\kbrr_{XZYZYZYZ}$	&$	-0.3	\pm	0.3	$	\\
\hline
\hline
\end{tabular}
\caption{\label{k2results}
Constraints 
(2$\si$, units $10^{-9}$~m$^{2}$)
on 72 independent coefficients $\krr_{\abgd\klmn}$
taken one at a time.}
\end{table*}

Using Eqs.\ \rf{pot} and \rf{k},
the Lorentz-violating acceleration at any position due to the source 
can be obtained via an integral over the geometry of the source mass.
The Lorentz-violating signal between the source and test masses
can then be extracted
by a further integration over the geometry of the test mass. 
The data are fitted to a Fourier series in the sidereal time $T_\oplus$,
\bea
\ta _{\rm LV}(T_\oplus) &=& 
C_0 + \sum\limits_{m = 1}^4 
\fr{{\sin (m{\om _ \oplus }\De T/2)}}{{m{\om _ \oplus }\De T/2}}
\nonumber\\
&&
\times
[ C_m \cos (m \om _\oplus T_\oplus) 
+ S_m \sin (m \om _\oplus T_\oplus) ].
\qquad
\label{fit}
\eea
In this expression,
the Fourier amplitudes are functions of 
the test mass and source mass geometry, 
the laboratory colatitude,
and the coefficients $(\kb_{\rm eff})_{JKLM}$
in the Sun-centered frame.
For a sinusoidal signal at frequency $m{\omega _ \oplus }$, 
the data average over  $\De T$ intervals 
leads to an attenuation of the amplitude 
by the factor 
$1 - {\sin (m{\omega _ \oplus }\Delta T/2)}/{m{\omega _ \oplus }\Delta T/2}$, 
which is approximately zero. 

Table \ref{tab1} displays the measurements
of the nine signal Fourier components.
The results for the HUST-2015 experiment
extracted from the fit \rf{fit}
are shown in the second column.
The third column contains
the results from the HUST-2011 analysis
\cite{hust15},
while the fourth and fifth columns
show those from the IU-2012 and IU-2002 analyses. 
Note that the HUST data are torque amplitudes 
while the IU data are force amplitudes,
with similar relative errors.

The HUST-2015 apparatus was designed 
to detect a non-Newton force at $8f_0$,
for which the Newton force is compensated by the design.
However,
the present work uses the data at frequency $16f_0$,
for which the Newton force is imperfectly compensated.
The errors arise from uncertainties in the dimensions and locations 
of the test and source masses.
This leads to a comparatively large systematic error 
that is restricted to the DC Fourier component $C_0$ of the $16 f_0$ signal
and can be seen in the lower panel of Fig.\ \ref{dat}.
In general,
experiments searching for sidereal signals are susceptible
to systematic errors arising from mundane diurnal variations.
For HUST-2015,
the dominant contribution of this type 
is due to temperature fluctuations.
This affects the torque in two main ways. 
First, 
it changes the dimensions and relative positions 
of the pendulum and the attractor,
which leads to a variation of the amplitude of the $16 f_0$ Newtonian torque. 
The temperature is recorded synchronously throughout the data collection. 
For a typical temperature dataset 
taken over a period of 9.5 days in March 2015, 
only diurnal and semidiurnal variations 
with amplitudes of $3.8\pm 1.3$ mK and $4.9\pm 0.7$ mK are evident. 
The largest variation of the geometric parameters 
is the relative height between the pendulum and the attractor, 
which is about 0.04 $\mu$m, 
resulting in a negligible variation of $<0.006\times 10^{-16}$ Nm
in the amplitude of the $16 f_0$ Newtonian torque.
Second, 
the temperature fluctuation changes the equilibrium positon 
of the torsion balance.
This leads directly to a torque variation, 
with a temperature-to-torque coefficient 
of $(1.1 \pm 0.2)\times 10^{-12}$ Nm/K. 
However,
the relevant concern is the diurnal and harmonic variations 
of the amplitude of the $16 f_0$ signal,
which are $<(2\pm 2)$ $\mu$K,
resulting in a torque noise of $<0.06\times 10^{-16}$ Nm 
at the $2\si$ level. 
This is included in the statistical errors listed in Table I. 
In effect,
the total noise is stable in the frequency band relevant 
to the Lorentz-violating signal,
so the uncertainties for the modes $C_m$ and $S_m$
are the same size and dominated by statistics. 
The net results of this error budgeting are shown in Table \ref{tab1}.

Simultaneous analysis of all these data
yields independent measurements of 14 accessible degrees of freedom.
Table \ref{combined} displays these measurements
in the Sun-centered frame.
A convenient choice of the 14 independent effective coefficients
has been made,
and the table lists them alphabetically by indices.
The 14 independent effective coefficients $(\kb_{\rm eff})_{JKLM}$ 
appearing in Table \ref{combined}
are linear combinations of a subset of the fundamental coefficients 
$\kddr_{\abgd\kl}$ and $\krr_{\abgd\klmn}$
that govern quadratic curvature derivatives and couplings
in Lorentz-violating gravity
\cite{bkx}.
The relationship is 
\bea
(\kb_{\rm eff})_{JKLM} &=& 
-2 \kbddr_{T(JTKLM)} -2 \kbddr_{N(JKNLM)}
\nonumber\\
&&
\hskip-25pt
- 2 \kbddr_{NTNT(JK} \de_{LM)} - \kbddr_{NPNP(JK} \de_{LM)}
\nonumber\\
&&
\hskip-25pt
+8 \kbrr_{T(JTKTLTM)} + 8 \kbrr_{N(JNKPLPM)} 
\nonumber\\
&&
\hskip-25pt
+ 16 \kbrr_{T(JTKNLNM)} ,
\label{keffpap}
\eea
where the symmetry indicated by the parentheses 
is on the spatial indices $JKLM$ only,
and $T$ is the temporal index.
The parameter $a$ used in Eq.\ (6) of Ref.\ \cite{bkx} has been set to zero,
which is possible without loss of generality
\cite{km16}.
Counting coefficients reveals that
Eq.\ \rf{keffpap} involves
63 of the 126 independent degrees of freedom in
$\kddr_{\abgd\kl}$ and 
78 of the 210 independent degrees of freedom in
$\krr_{\abgd\klmn}$.
However,
some of these are rotation invariants
and hence cannot be detected via sidereal studies,
implying that our analysis achieves sensitivity to 
59 independent degrees of freedom in $\kddr_{\abgd\kl}$ 
and 72 in $\krr_{\abgd\klmn}$.
Using Eq.\ \rf{keffpap},
we can transform the 14 independent measurements
given in Table \ref{combined}
into limits on a conveniently chosen set 
of $59+72=131$ independent coefficients
taken one at a time,
following standard procedure in the field
\cite{tables}.
This procedure yields the measurements 
shown in Tables \ref{k1results} and \ref{k2results}.

To summarize, 
a combined analysis of data from the short-range experiments 
HUST-2015, HUST-2011, IU-2012, and IU-2002
constrains simultaneously and independently 
14 combinations of coefficients for Lorentz violation,
consistent with no effect at the level of $10^{-9}$ m$^{2}$.
This represents an improvement of an order of magnitude
over previous experimental analyses,
achieving sensitivity to 131 independent types of  
Lorentz-violating quadratic curvature derivatives and couplings.
The results presented in this work complement
recent limits obtained on 25 independent coefficients 
\cite{kt15}
from gravitational \v Cerenkov radiation 
and on 39 independent coefficients
\cite{km16}
from the gravitational-wave event GW150914
\cite{ligo}.
They also complement experimental searches 
for a Lorentz-violating inverse-{\it square} law
\cite{2007Battat,2007MullerInterf,2009Chung,%
jcl10,2012Iorio,2013Bailey,2014Shao,he15,bk,bo16}.

This work was supported by 
the National Natural Science Foundation of China 
(11275075, 11325523, and 91436212) 
and 111 project (B14030), 
by the Australian Research Council Grant DP160100253,
by the United States National Science Foundation 
under grants PHY-1207656 and PHY-1402890,
by the United States Department of Energy
under grant {DE}-SC0010120,
and by the Indiana University Center for Spacetime Symmetries.

\end{document}